\documentstyle[prl,aps,floats,epsf]{revtex}
\begin{document}
\draft
\tighten
\twocolumn[\hsize\textwidth\columnwidth\hsize\csname
@twocolumnfalse\endcsname
\draft 
\title{
Carrier Density Collapse and Colossal Magnetoresistance
in Doped Manganites
}
\author{ A.S. Alexandrov$^{1,*}$ and A.M. Bratkovsky$^{2,\dagger}$}

\address{$^1$Department of Physics, Loughborough University,
Loughborough LE11 3TU, UK\\
$^2$Hewlett-Packard Laboratories, 3500~Deer~Creek
Road, Palo Alto, California 94304-1392 }
\date{May 28, 1998}
\maketitle
\begin{abstract}
A novel  ferromagnetic transition,
accompanied by carrier density collapse,
is found in  doped charge-transfer
insulators with strong electron-phonon coupling.
 The transition is driven by an exchange interaction of
polaronic carriers with localized  spins;
the strength of the interaction determines whether the
transition is first or second order.
A giant drop in the number of current carriers during the transition,
which  is a consequence of bound pairs
formation in the paramagnetic phase close to the transition,
is extremely sensitive to an external
magnetic field. This carrier density collapse
describes  the resistivity peak and the colossal
magnetoresistance  of  doped  manganites.
\end{abstract}
\vskip 2pc ] 
\narrowtext

The interplay of the electron-phonon and
exchange interactions\cite{alemot,bis,fes}
is thought to be responsible for many exotic properties of oxides
ranging from high-$T_{c}$ superconductivity in cuprates \cite{bed} to
colossal magnetoresistance
(CMR) and ferromagnetism in doped manganites \cite{van,jin,sch,shi,ram}.
A huge negative magnetoresistance was observed in 
doped perovskite manganites
La$_{1-x}$D$_x$MnO$_3$ (D=Ca, Sr, Ba)
close to the ferromagnetic transition in a certain range of
doping $x \approx 0.15-0.4$ \cite{jin,sch,shi,ram},
and this raised a question of possible applications.

The metal-insulator transition in lanthanum manganites has long been
thought to be the consequence of a double exchange (DEX), which results in
a varying band width of holes doped into the Mn$^{3+}$ $d$-shell
\cite{dou}, as function of the doping concentration and temperature.
Recently it has been  realized \cite{mil}, however, that the effective
carrier-spin 
interaction in DEX model is too weak to lead
to a  significant reduction of the electron band width
and, therefore, cannot account for  the observed scattering rate \cite{edw}
(see also Ref. \cite{fis}) or for localization induced
by slowly fluctuating spin configurations \cite{var}.
In view of this problem, it has
been suggested \cite{mil} that the essential physics of perovskite
manganites lies in the strong coupling of carriers to
Jahn-Teller lattice distortions.
The argument \cite{mil} was that
in the high-temperature state the electron-phonon coupling
constant $\lambda$ is large
(so that the carriers are  polarons)
while the growing ferromagnetic order
increases the bandwidth and thus decreases $\lambda$
sufficiently for  metallic behavior to set in  below
the Curie temperature $T_{c}$.
A giant isotope effect \cite{mul},
the sign anomaly of the Hall effect,
and the Arrhenius behavior of the drift  and Hall mobilities \cite{emin}
over a temperature range from $2T_{c}$ to $4T_{c}$ unambiguously confirmed
the polaronic nature of carriers in manganites.

However, an early established unusual relation between magnetization and
transport below $T_c$
have led to a conclusion that the polaronic hopping is the prevalent
conduction mechanism also below $T_c$\cite{hundley}.
Low-temperature optical \cite{oki1,oki2,kim},
electron-energy-loss
(EELS)\cite{ju} and photoemission spectroscopies \cite{des}
showed
that the idea \cite{mil,var}
of a `metalization' of manganites below $T_{c}$  is not
tenable.
A broad incoherent spectral feature \cite{oki1,oki2,kim,des} and a pseudogap
in the excitation spectrum \cite{des,wei} were observed while
the coherent Drude weight
appeared to be {\em two orders} of magnitude smaller \cite{oki2} than
is expected for a metal.
EELS \cite{ju} confirmed that  manganites
are charge-transfer type doped insulators
having $p$-holes as the current carriers rather than $d$ (Mn$^{3+}$)
electrons.
The photoemission and O $1s$ x-ray absorption spectroscopy of
La$_{1-x}$Sr$_x$MnO$_3$ showed that the
itinerant holes doped into LaMnO$_3$ are indeed of oxygen $p$
character, and their coupling with $d^4$ local moments on Mn$^{3+}$ ions
aligns the moments ferromagnetically \cite{saitoh}.
Moreover, measurements of the  mobility \cite{ram,htkim} do not
show any field dependence.
The calculated resistivity
 is in poor agreement with the data  and the characteristic
theoretical
field ($\sim 15$T) for CMR is too high
compared with the experimental one ($\sim 4$T)\cite{mil}.
As a result, self-trapping  above $T_{c}$
and the idea of metalization below $T_{c}$
do not explain  CMR either.
Carriers retain their polaronic
character well below $T_{c}$, as  manifested also
in the measurements of resistivity and thermoelectric power
under pressure \cite{gud}.

In the present paper, we propose a theory
of the ferromagnetic/paramagnetic phase transition 
in doped charge-transfer magnetic insulators
accompanied by
a  current carrier density collapse (CCDC)  and  CMR.
Taking into account a tendency of polarons to form bound pairs and
the (competing with binding) exchange
interaction of $p$ polaronic holes with $d$ electrons,  we find
a novel ferromagnetic transition driven by non-degenerate polarons. 
As a result we describe
the magnetization and
temperature/field dependence of the resistivity
of La$_{1-x}$Ca$_{x}$MnO$_{3}$  close to $T_{c}$ in a region $0.15<x<0.4$.


The Hamiltonian containing the physics compatible
with  the experimental observations is
\begin{eqnarray}
{\cal H}&=&\sum_{k,s} E_{ {\bf k}} h^{\dagger}_{{\bf k}s} h_{{\bf k}s}
 - \frac{J_{pd}}{2N} \sum_{{\bf k},i}m_{\bf k}S^z_i
 +{\cal H}_{\rm Hund}\nonumber\\
&+&(2N)^{-1/2}\sum_{{\bf k,q},s}\hbar\omega_{\bf q}
\gamma_{\bf q} h^{\dagger}_{{\bf k+q}s} h_{{\bf k}s}
(b_{\bf q}  -  b^\dagger_{\bf -q})\nonumber\\
&+&\sum_{\bf q} \hbar\omega_{\bf q}
(b^\dagger_{\bf q}b_{\bf q} + 1/2),
\label{eq:ham}
\end{eqnarray}
where $E_{ {\bf k}}$ is  the LDA energy dispersion \cite{pic},
$ h_{ {\bf k}s}$ is
the annihilation hole operator
 of a (degenerate) $p$ oxygen band with spins
$s=\uparrow$ and $\downarrow$,
$J_{pd}$ is the exchange
interaction of $p$ holes with four $d$ electrons of
the Mn$^{3+}$ ion at the site $i$, $m_{{\bf k}}\equiv
h^{\dagger}_{ {\bf k}\uparrow} h_{{\bf k}\uparrow}-
h^{\dagger}_{ {\bf k}\downarrow} h_{{\bf k}\downarrow}$,
$S^z_{i}$ is the $z$-component of Mn$^{3+}$ spin,
which is $S=2$ due to the
strong Hund coupling, ${\cal H}_{\rm Hund}$,
of the four $d$-electrons on Mn$^{3+}$ sites,
$N$ is the number of unit cells.
The two last terms of the Hamiltonian describe the
coupling of $p$ holes  with phonons and the phonon energy,
respectively ($\omega_{\bf q}$ is the phonon frequency, 
$\gamma_{\bf q} = -\gamma_{\bf -q}^*$ is the coupling constant \cite{alemot}).
 If the holes
were doped into $d$ shell instead of $p$ shell, the Hamiltonian
would be similar to the Holstein $t$-$J$ model \cite{fes}
with about the same physics of CMR as proposed below.


Essential results are readily obtained
within the Hartree\--Fock approach
for the exchange interaction \cite{boo} and the Lang-Firsov polaron
trans\-formation, which eliminates terms linear in the electron-phonon
interaction in Eq.~(\ref{eq:ham})\cite{alemot}.
Thus, we find spin-polarized $p$ bands
\begin{equation}
\epsilon_{ \bf k\uparrow(\downarrow)}
=\epsilon_{\bf k}  -(+) \frac{1}{2}J_{pd} S \sigma -(+)\mu_{B}H,
\label{eq:pol}
\end{equation}
where $\epsilon_{\bf k} = E_{\bf k} e^{-g^2}$, and
$g^2 \sim \gamma^2$ describes the polaronic band narrowing
\cite{alemot} and the isotope effect \cite{mul}, $\sigma$ is the
normalized thermal average of the Mn spin (\ref{eq:sigma});
$H$ is the external magnetic field, and $\mu_{B}$ is the Bohr magneton.
The $p-d$ exchange interaction depends only on total (average)
magnetization because we
assume that the system is homogeneous. 
The magnetization of Mn$^{3+}$ ions is given by
\begin{equation}
\sigma\equiv  \langle S^z_i\rangle/S
=B_{S}\left({J_{pd}m + 2g_{\rm Mn}\mu_{B} H \over{2 k_B T}}\right),
\label{eq:sigma}
\end{equation}
with $m$ the absolute value of the magnetization of holes determined as
\begin{equation}
m \equiv  {1\over N} \sum_{\bf k}\langle m_{\bf k}\rangle
=\int d\epsilon N_{\rm p}(\epsilon) \left[ f_{\rm p}(\epsilon_{\bf k\uparrow})
-f_{\rm p}(\epsilon_{\bf k\downarrow})\right].
\end{equation}
Here $B_{S}(x) =\- [1+1/(2S)]\- \coth[(S+1/2)x]\- -[1/(2S)]\coth(x/2)$
is the Brillouin function, $g_{\rm Mn}$ the Lande $g$-factor for
Mn$^{3+}$ in a manganite,
$N_{\rm p}(\epsilon)$  the density of states in the narrow polaron band,
and $f_{\rm p}(\epsilon_{{\bf k}s})=[y^{-1}\exp(\epsilon_{{\bf
k}s}/k_{B}T)+1]^{-1}$ 
the Fermi-Dirac distribution function
with $y=\exp(\mu/k_B T)$ determined by the chemical potential $\mu$.
Note that for $J_{pd}<0$ (antiferromagnetic coupling)
the main system of equations (\ref{eq:n})-(\ref{eq:sig}) remains the
same after a substitution $J_{pd} \rightarrow |J_{pd}|$.

Along with the band narrowing effect,
the strong electron-phonon interaction
binds two holes into a pair
 (bipolaron) \cite{alemot}.
The bipolarons are practically
immobile in cubic manganites because
the electron-phonon interaction is too strong
in contrast with cuprates, where
 bipolarons are mobile  owing to their geometry  and
a moderate coupling with phonons\cite{chak}.

If these bound pairs are extremely local objects, two holes on the same
oxygen, then they will form singlet. If, however, these holes are
localized on different oxygens, then they may 
form a triplet state.
Because of their zero spin, the only role of singlet bipolarons in manganites
is to determine the chemical potential $\mu$, which can be found
with the use of the
total carrier density per cell $x$ as
$
2\int dE N_{\rm bp}(E) f_{\rm bp}(E)=x-n,
$
where $N_{\rm bp}(E)$  the density of bipolaronic states,
$f_{\rm bp}(E)=\{y^{-2}\exp[(E-\Delta)/k_{B}T]-1\}^{-1}$  the bipolaron
distribution function, $\Delta$ the bipolaron binding energy, $n$ is
the density of single (unbound) hole polarons, which are the only current
carriers in manganites, and $x$ is the doping concentration.

It is the localization of $p$-holes into  immobile bound pairs
combined with their exchange interaction
with the Mn $d^4$ local moments
that are responsible for CMR. The density of these pairs has a sharp
peak at a ferromagnetic transition when system is cooled down through
the critical temperature $T_c$. Below $T_c$ the binding
of polarons into pairs competes with the ferromagnetic exchange
which tends to align the polaron moments and, therefore, breaks
those pairs apart. These competing interactions lead to unusual
behavior of CMR materials and a huge sensitivity of their transport to
external field.

To illustrate the point we assume that $T_{c}$
is comparable with the polaron, $W$, and bipolaron band widths
\cite{twfac}. 
Then (bi)polarons are not degenerate in the relevant temperature range,
$f_{\rm p}\simeq y\exp(-E/k_{B}T)$ and $f_{\rm bp}\simeq y^2
\exp[(\Delta-E)/k_{B}T]$,
and we can evaluate integrals
reducing the system of mean field equations to
\begin{eqnarray}
n&=& 2\nu y  \cosh[(\sigma+h)/ t],\label{eq:n}\\
m&=&n \tanh[(\sigma+h)/ t],\label{eq:m}\\
\sigma&=&B_2 [(m+4h)/(2 t)],\label{eq:sig}
\end{eqnarray}
and
\begin{equation}
y^{2}= {x-n\over{2\nu^2}} \exp(-2\delta/t).
\label{eq:y}
\end{equation}
Here we use the dimensionless temperature $t=2k_{B}T/(J_{pd}S)$,
magnetic field $h=2\mu_{B}H/(J_{pd}S)$, and the binding energy
$\delta\equiv \Delta/(J_{pd}S)$, while
$\nu(=3)$ is the degeneracy of the $p$ band.

The  polaron density $n$ is determined by Eq.~(\ref{eq:n})
with $\sigma=0$ above $T_{c}$.
At the critical temperature, the polaron density has a minimal value
$n_{c}\simeq (2x)^{1/2}\exp(-\delta/t_{c})$, it then grows
exponentially with temperature and saturates
at $n=(1+2x)^{1/2}-1$.
The remarkable  observation
is that there is a sharp increase of the polaron density
at  temperatures below $T_{c}$.
The physical origin of the unusual minimum of
the current carrier density at $T_{c}$ lies
in the instability of bipolarons below $T_{c}$ due to the exchange
interaction of polarons with $d$ electrons.
The spin-polarized polaron band
falls below the bipolaron band with decreasing temperature, so that
all carriers are unpaired at $T=0$ if $J_{pd}S\geq \Delta$.

Linearizing (\ref{eq:n})-(\ref{eq:sig})
we find the critical temperature in zero magnetic field is
$
t_{c}= (n_{c}/2)^{1/2},
$
where the  polaron density at the transition $n_{c}$ is determined by
\begin{equation}
n_c^{1/2} \ln{2(x-n_c)\over{n_c^2}}=2^{3/2}\delta.
\end{equation}
This equation has  solutions only for $\delta$ below some critical value
$\delta_{c}(x)$, Fig.~1(b, inset). 
The numerical solution
of the system Eqs.~(\ref{eq:n})-(\ref{eq:y})
shows that for $\delta>\delta_{c}(x)$
the ferromagnetic phase transition is first order
with jumps of the polaron density and the magnetization,
as observed  \cite{kuw},  Fig.~1(a).
The transition is continuous when $\delta<\delta_{c}(x)$, Fig.~1(b).
\vspace{.75in}
\begin{figure}[\!h]
\epsfxsize=3.4in
\epsffile{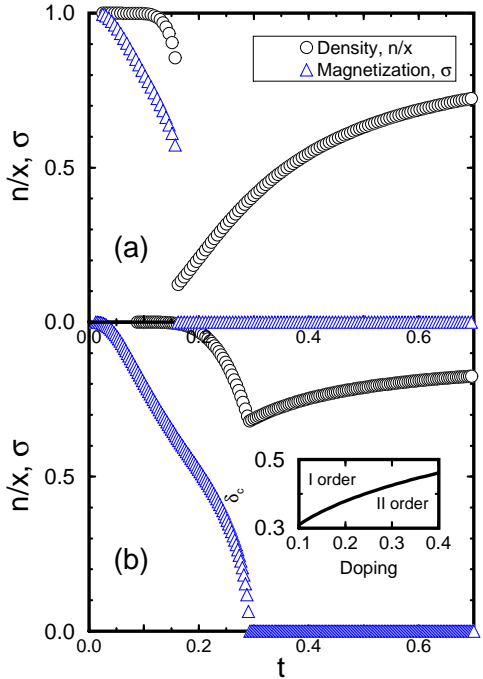 }
\caption{
Polaron density $n/x$ and magnetization as a function of  temperature
$t=2k_{B}T/(J_{pd}S)$ 
in a doped charge-transfer insulator
near (a) the first order,
$\delta\equiv\Delta/(J_{pd}S)=0.5$, and (b) second order,
$\delta=0.25$,
phase transitions (doping $x=0.25$). Inset: the critical value of the
relative binding energy of polaron pairs $\delta$ separating the regions of
the first and second order  phase transitions.
\label{fig:1}}
\end{figure}
A relatively weak magnetic field has a drastic effect on the inverse
carrier density, $1/n$, near the
first order transition, as shown in Fig.~2.
As a result, the resistivity $\rho=1/(en \mu_{\rm p})$
has a sharp maximum, which is
extremely sensitive to the magnetic field in the vicinity of $T_{c}$.

It is assumed, as is usually the case, that the triplet states always
lie higher in energy than the singlet state. 
If the singlet-triplet separation becomes smaller than
the gap, $J_{st} \lesssim \Delta$,  then, because of a higher
number of the triplet states, their thermal population leads to a
deeper minimum in the density of polarons. We make an essential
assumption that the exchange between spins on Mn and triplet bipolarons is
suppressed because the bipolarons are strongly localized. Otherwise, the
triplet bound pairs, if they were formed in the paramagnetic phase, can
survive in the ferromagnetic phase thus reducing the carrier density
collapse. 
One can draw an analogy of this situation with singlet magnetism,
e.g. in Pr compounds \cite{singmag}.

In fact, our theory, Eqs.~(\ref{eq:n})-(\ref{eq:y}),
describes all the major features of the
temperature/field dependence of $\rho(T)$ \cite{sch}, with
a temperature and  field independent polaron drift mobility
$\mu_{\rm p}$\cite{mup}
in the experimental range of the magnetic field, Fig.~3.
\begin{figure}[t]
\epsfxsize=3in
\epsffile{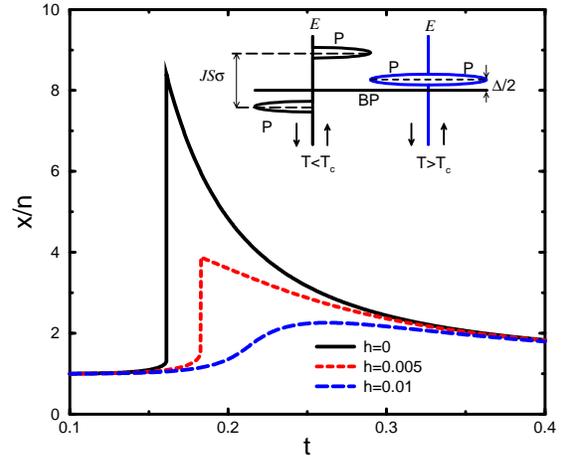 }
\caption{Inverse polaron density for different magnetic fields,
$\Delta/J_{pd}S=0.5$, doping $x=0.25$.
Note that the transition is a strong first order, and becomes
continuous only when the external magnetic field exceeds some critical
value. Inset: schematic of polaron (P) and bipolaron (BP) densities of
states at temperatures below and above $T_c$ for up ($\uparrow$) and
down ($\downarrow$) spin moments. The pairs (BP) break below $T_c$
if exchange $J_{pd}S$ exceeds the pair binding energy $\Delta$.
\label{fig:2}
}
\end{figure}
\begin{figure}[\!h]
\epsfxsize=3.4in
\epsffile{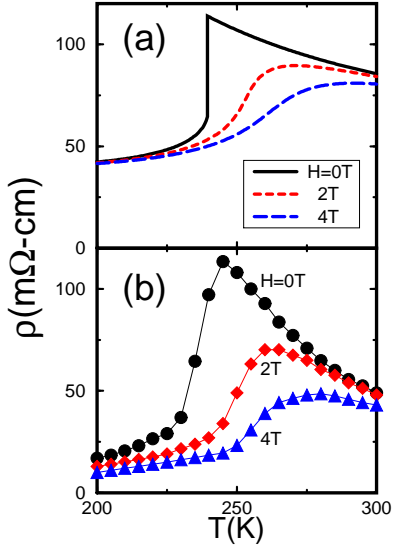 }
\caption{
Resistivity of La$_{0.75}$Ca$_{0.25}$MnO$_{3}$
calculated within the present theory
for the pair binding energy $\Delta=900$K, and polaron exchange
with the localized Mn$^{3+}$ spins
$J_{pd}S=2250$K (a), compared with experiment [7]
on panel (b).  Note an extreme sensitivity
of the theoretical resistivity to external magnetic field (a), observed
experimentally (b) for the doped manganite.
\label{fig:3}
}
\end{figure}
That suggests that CCDC is the origin of CMR.
In general, one has to take into account the temperature dependence of the
polaron mobility to extend our theory for temperatures far away from
the transition.

We have also compared this scenario with
the localization of $p$-holes due to a random
field with a
gap $\Delta/2$ between  localized impurity levels
and the conduction band \cite{alebra2}.
 We have found
the same features of the phase transition in zero field.
However, Zeeman  splitting of the impurity states,
and a different  behavior of $y$ with
density, makes the transition {\em far less
sensitive to the magnetic field}. As a result, {\em no}
quantitative description of the experimental
CMR  has been found in this case.

In conclusion, we have  found that  a few non-degenerate
polarons in the $p$ band   polarize localized $d$ electrons 
because of a huge density of states in the narrow polaronic band. For
a sufficiently large $p-d$ exchange $J_{pd}S > \Delta$,  we have
obtained current carrier density collapse  at the transition owing to
the formation of immobile bipolarons in the paramagnetic phase with
the binding energy $\Delta$ x\cite{alemot}. 
Competition between the binding energy of polarons, which promotes a
formation of local pairs, and their exchange interaction
with $d$ electrons, which breaks
them at lower temperatures, results in a huge negative
magnetoresistance close to the ferromagnetic transition.

We have explained the resistivity peak and the colossal
magnetoresistance  of  doped  perovskite manganites, Fig.~3, as a
result of the current carrier density collapse.
Depending on the ratio $\Delta/(J_{pd}S)$, the transition is  first
 or second order, Fig.~1. 
It is not clear at present whether the main idea underlying this
picture, the assumption of singlet bound states of charge carriers, is
true for manganites. The available experimental data, e.g.
the  tunneling spectroscopy \cite{wei} suggests that it is.
Our goal is to stimulate a wider
discussion and new experiments in this direction.
We expect that the present theory is general enough to also account for
the giant magnetoresistance observed in pyrochlore manganites \cite{shi}.

We acknowledge useful discussions with A.R.~Bishop, G.A.D.~Briggs,
D.M.~Edwards, M.F.~Hundley, 
P.B.~Littlewood, S.~von~Molnar, V.G.~Orlov, W.E.~Pickett, D.J.~Singh,
S.A.~Trugman, and R.S.~Williams. We thank G.~Aeppli, D.S.~Dessau,
H.T.~Kim,  A.P.~Ramirez, and G.-m. Zhao for the information on their data
and helpful discussions.




\begin{references}
\bibitem{twfac} Otherwise
there will be an inessential factor $(T/W)^{d/2}$ in Eq.~(\ref{eq:n})
for $n$, where $d$ is the dimensionality of a system.
\bibitem{mup} With our low polaron density at the transition
 the polaron mobility is
$\mu_{\rm p}=0.2$~cm$^2$/Vs
for $x=0.33$ \cite{emin}, and about $0.03$~cm$^2$/Vs for $x=0.25$
\cite{sch} that
lies in the range typical of polaronic conductors like TiO$_2$ at room
temperature \cite{alemot}.
\end{references}
\end{document}